\documentstyle[12pt,aasms4]{article}
\include{psfig}

\begin{document}
 
\title{GALAXY MORPHOLOGY WITHOUT CLASSIFICATION~: SELF ORGANIZING MAPS}

\author{Avi Naim, Kavan U. Ratnatunga and Richard E. Griffiths}

\affil{The Johns Hopkins University, Department of Physics \& Astronomy, \\
Baltimore, MD 21218, U.S.A.~$^1$}
 
\bigskip

$^1$ current address : Department of Physics, Wean Hall, \\
Carnegie Mellon University, 5000 Forbes Ave., \\
Pittsburgh, PA 15213, U.S.A.

\begin{abstract}

We examine a general framework for visualizing datasets of high ($> 2$) 
dimensionality, and demonstrate it using the morphology of galaxies at 
moderate redshifts. The distributions of various populations of such
galaxies are examined in a space spanned by four purely morphological 
parameters. Galaxy images are taken from the Hubble Space Telescope (HST) 
Wide Field Planetary Camera 2 (WFPC2) in the I band (F814W). Since we have
little prior knowledge on how galaxies are distributed in morphology space we
use an unsupervised learning method (a variant of Kohonen's Self Organizing
Maps, or SOMs). This method allows the data to organize themselves onto a 
two-dimensional space while conserving most of the topology of the original 
space. It thus enables us to visualize the distribution of galaxies
and study it more easily. The process is fully automated, does not rely on any
kind of eyeball classification and is readily applicable to large numbers of 
images. We apply it to a sample of 2934 galaxies, and find that morphology 
correlates well with the apparent magnitude distribution and to lesser extents 
with color and bulge dominance. The resulting map traces a morphological 
sequence similar to the Hubble Sequence, albeit two dimensional. We use the
SOM as a diagnostic tool, and rediscover a population of bulge-dominated
galaxies with morphologies characteristic of peculiar galaxies. This is
achieved {\it without recourse to eyeball classification}. We also examine
the effect of noise on the resulting SOM, and conclude that down to I
magnitude of 24 our results are reliable. We propose using this method as a   
framework into which more physical data can be incorporated when they 
become available. Hopefully, this will lead to a deeper understanding of 
galaxy evolution.

\end{abstract}

\keywords {galaxies: morphology - galaxies: evolution - galaxies: peculiar}

\section{Introduction}

Morphological classification of galaxies was originally envisaged as a tool
for studying the evolution of galaxies (e.g., Hubble 1936). Much like other
fields of science, as the amount of data grew the classifications were revised
and became more and more refined (Sandage 1961; de Vaucouleurs 1959; van den
Bergh 1960, 1976). At some point the question arose as to how well those
refinements correlate with physical quantities and processes within galaxies. 
In an excellent review, Roberts \& Haynes (1994) show that morphological types 
in the local universe do correlate with color, HI mass  and other quantities 
{\it in the mean}, but there is a large scatter about the mean. This implies 
that morphological classification has become overly refined, at least as far as
its relation to physical properties is concerned.

A major limitation of most classification schemes for galaxies is that they 
were devised solely using samples of nearby galaxies, due to the lack of 
imaging capabilities at higher redshifts. This situation has changed with the 
advent of the Hubble Space Telescope and very large ground based telescopes. 
The morphology of large numbers of galaxies at moderate redshifts ($z < 1$) is 
now available, and preliminary results (Griffiths {\it et al.} 1994; 
Glazebrook {\it et al.} 1995; Driver {\it et al.} 1995; Abraham {\it et al.}
1996) indicate that many galaxies at moderate redshifts do not fit comfortably
on the Hubble sequence. It is an obvious challenge to try and incorporate 
galaxies at different redshifts into one coherent scheme.

A lot of work has been done recently on morphological classification of faint 
galaxy images. Most of it, however, relies on eyeball classifications : Cowie 
{\it et al.} (1995) present deep I band WFPC2 images of a K selected sample. 
They give a qualitative, eyeball account of the change they see in the 
morphology of galaxies around $K = 19.5$. Driver {\it et al.} (1995) divide 
galaxies in a deep WFPC2 field into three eyeball classes and analyse the 
number counts as a function of type. van den Bergh {it et al.} 
(1996) produced a morphological catalogue of galaxies in the Hubble Deep
Field (HDF), which was again based on eyeball classifications. In addition, 
they supply two quantitative parameters for those galaxies (light concentration
and asymmetry), which allow for a more objective analysis. Odewahn {\it et 
al.} (1996) use both eyeball classifications and trained artificial neural
networks to obtain classifications for galaxies in deep HST fields. Their
network utilizes parameters derived from surface brightness profiles in U, B, 
V and I filters. The move from pure eyeball classification to automated
classification using objective parameters has been inevitable, due to the
large quantities of images that have become available over the past few years.
The parameters used by van den Bergh {\it et al.} (1996) proved to be a useful
first step in this direction, although they gave a very crude separation of
eyeball types. Using light profile parameters Odewahn {\it et al.} (1996) 
discuss the possible makeup of the population of blue galaxies. Both of these 
papers tie their quantitative parameters to classifications on the existing 
Hubble sequence, which is apparently insufficient for the full range of 
morphologies detected with HST. 

Since the Hubble Sequence appears too refined on the one hand, and not general
enough on the other, we suggest a more general approach here. In recognition of
the fact that morphology is a continuous quantity we abandon any attempt to
tag each galaxy with a specific type. Instead, we use a space spanned by four
morphological parameters, which was introduced elsewhere (Naim {\it et al.} 
1997), and examine the distribution of various populations of galaxies in it. 
We start with a large, complete, magnitude-limited sample of HST Wide Field 
Planetary Camera 2 (WFPC2) images, which is described in \S~2. We have little
prior knowledge of the distributions of galaxies in this space. For this reason
we use a variant of an {\it unsupervised learning} technique called Self
Organizing Maps (SOMs). It allows data taken from a space of high 
dimensionality to organize themselves into a two dimensional "histogram" while 
retaining most of the topology. The resulting map can then be plotted and 
analyzed. SOMs, which are explained in detail in \S~3, therefore combine 
non-linear clustering with a dimension-reduction technique. SOMs have been 
little used in astronomy to date (the one example we are aware of is 
M\"{a}h\"{o}nen \& Hakala 1995) and, as we show below, prove a valuable tool 
for unsupervised data analysis. However, one important point has to be stressed
from the outset : we are using a non-parametric method here, in the sense that 
the results are not described in terms of functional dependencies between the
parameters we use. Consequently, the SOMs are primarily a {\it diagnostic 
tool}, which should be used only as a first step towards forming a model that 
explains the observations. Its most important feature is the ability to 
identify special populations that merit closer examination. We first 
demonstrate the application of SOMs to a synthetic dataset (\S~4) and then 
apply them to the sample of HST galaxies (\S~5). The discussion follows in 
\S~6.

\section{Sample Selection and Morphological Parameters}

\subsection{Sample Selection}

It is easiest to select a suitable, large sample from data that were collected 
uniformly. The 27 contiguous fields of the Groth-Westphal Strip (Groth {\it et 
al.} 1995) make an excellent such collection. I band (F814W) images were 
preferred over V band (F606W) images (which are also available for the same 
fields) for two reasons : first, exposures in I were about $50\%$ longer and 
typically resulted in higher signal-to-noise ratio images; Second, at the 
expected redshifts of these galaxies the I filter corresponds roughly to the 
rest frame B band, for which most existing morphological schemes were defined, 
while the V filter corresponds to a much bluer rest frame band in which images 
appear much more broken up. 

Our indications from previous work (Naim {\it et al.} 1997) are that down to
an isophotal magnitude of $I = 24.0$ a distinction between morphologically 
``normal'' and ``peculiar'' galaxies is still possible, although it suffers 
increasingly from effects of noise towards the faint end. We decided to go for 
the same limit here and then examine a higher signal-to-noise subset of the 
sample, to see what effect the noise had on our results. There were 3391 images
brighter than $I = 24$ in the Groth-Westphal strip. The MDS pipeline, using a
maximum likelihood method (Ratnatunga {\it et al.} 1997), fits simple 
photometric models ($r^{1/4}$ law, exponential disk and combinations of the 
two) to galaxy images. It was found that the fitted half light radius parameter
is very useful in separating stars and compact objects from galaxies, and the 
limiting value was empirically set at $0.1$ arcsec (1 image pixel). It is clear
that some distant galaxies as well as closer compact objects have half light
radii smaller than this limit. Therefore, not all of the 421 images which
were removed from the sample due to failing this test are indeed stars. 
However, images whose half light radius is smaller than $0.1$ arcsec are
typically no more than 3-4 pixels across, thus containing almost no 
morphological information. Consequently, we use this cutoff not only as a 
safeguard against contamination by stars but also as a practical lower limit 
for the derivation of our parameters. On top of the 421 images mentioned above,
less than 20 other images were rejected by the program which calculates the 
morphological parameters, due to low quality (e.g., too high a fraction of 
missing pixels). During classifications by eye (see below) several more (less 
than 20) images were rejected due to other problems (e.g., a nearby star 
overlaps the galaxy). The final sample contains 2934 entries. 

Isophotal magnitudes are tightly correlated with the {\it integrated} 
signal-to-noise index, $\nu$, which is calculated by summing the individual
signal-to-noise ratios $> 1$ over image pixels. See Ratnatunga {\it et al.} 
1997 for details). Note that since it is the {\it integrated} signal-to-noise
the values we are dealing with are typically of order $100$. At the limiting 
magnitude of $I = 24.0$ all but six galaxies in the sample have $\nu > 100$, 
which is incidentally the limit below which no disk+bulge photometric model 
fit was attempted by the maximum-likelihood software (although pure bulge and 
pure disk models are attempted down to much lower values). 

\subsection{Morphological Parameters}

A full description of the four parameters we use was given in Naim {\it et al.}
(1997). We therefore give only a brief description of them here. In designing
these parameters we attempted to give as full a description as possible of the 
features that stand out in galaxy images, while remaining neutral with respect 
to quantities such as the underlying photometric model or the color of the 
image. Our parameters are :

\begin{enumerate}
\item{Blobbiness : The degree to which bright pixels stand out, accentuating
bright localized structure. This parameter may be related to regions of intense
star formation. Briefly, this parameter is calculated for each bright image 
pixel as the fraction of brighter pixels out of the total number of pixels in 
a semi-circular environment around it.}

\item{Isophotal Center Displacement : The displacement of geometrical centers
of various isophotes from each other, as a measure of overall asymmetry. This 
parameter may be related to merging histories by detecting tidal tails.}

\item{Isophotal Filling Factor : The fraction of pixels belonging to a certain
isophote out of the number of pixels in the ellipse enclosing that isophote.
This is a measure of overall structure : in featureless images this fraction
is expected to be higher than in images exhibiting a lot of structure, because
in the latter bright pixels will be found at higher radii, making the 
enveloping ellipse much bigger. This expectation is verified for an eyeballed
subset of our sample (Naim {\it et al.} 1997, see also below), in which late 
spirals and peculiars average a value of less than $0.2$ for this parameter, 
early spirals average close to $0.3$ and ellipticals/lenticulars average over 
$0.35$.}

\item{Skeleton Ratio of detected structures, indicating how elongated the 
structures are. Briefly, for detected structures in the galaxy image, this is 
the fraction of pixels making up the ``backbone'' of the structure to the 
total number of pixels in that structure.}
\end{enumerate}

The first three parameters are evaluated from the raw I band image while the
fourth is derived from the residual image, which is left after subtracting
the best fit photometric model (from the maximum-likelihood software).

\section{Self Organizing Maps}

The motivation behind self organizing maps (SOMs) derives from the inability
to plot data in more than three dimensions. Kohonen (1989) suggested a 
non-linear mapping from a given $M$-dimensional space ($M > 2$) onto a two
dimensional map, in a way that maintains as much as possible of the topology
of the higher dimension space. SOMs are therefore one implementation of {\it 
unsupervised learning}, a generic name referring to methods for describing data
without any prior knowledge of how they cluster. Self organization takes place
in an iterative manner with little user intervention. The role played by
the user is reduced to defining the organizing criterion (i.e., the criterion 
determining which vector is mapped to which node in the SOM). The resulting
map can be regarded as a two-dimensional histogram, although its axes do not
carry the usual parametric meaning. The numbers on the $x$ and $y$ axes 
represent positions in the map, not values of the $M$ parameters making up
the space of the data. 

Let a given dataset contain $N$ vectors of dimension $M$, each describing a 
single object (e.g., a galaxy). In the case of our galaxy sample, $N = 2934$
and $M = 4$. The ``data-space'' is therefore $M$-dimensional. Define the map 
as a two-dimensional array of discrete nodes. Throughout this paper we use
square maps of size $16$ nodes. The nodes occupy positions in what we refer to 
as the (two-dimensional) ``map space''. The link between the two spaces is 
realized by assigning each node of the map an $M$-dimensional 
``characteristic'' vector from the data-space. Note that this assignment is
done in an automated way, with no input from the user, i.e., it is truly an
unsupervised operation. The key measure in the process of self-organization is 
distance. Distances are calculated within each of the spaces independently. 
For better clarity we will refer to them in what follows as data-distance and 
map-distance, respectively. The user's role is confined to choosing a certain 
distance measure (e.g., the $L2$ norm, also known as the Euclidean distance), 
which serves as the organizing criterion. Each object in the dataset is mapped 
to the node whose characteristic vector is closest to it in the sense of that 
distance measure (the ``winning'' node). In each iteration of the training 
process the entire dataset is mapped to the SOM and then the characteristic 
vectors of the nodes are updated according to the objects mapped onto them. 
Topology is preserved by allowing nodes in the vicinity of the winning node to 
be updated as well. Over many iterations this will cause nodes which lie close 
to each other to obtain similar characteristic vectors, and therefore 
eventually whole regions in the SOM will correspond to specific populations in 
the dataset. While nearby nodes will represent finer details within each 
population, nodes far away from each other will represent significantly 
different populations. The iterations are stopped once some convergence 
criterion (see below) is met.

Normally one initializes the map nodes to have random characteristic vectors
at first. However, this could assign very different vectors to adjacent nodes,
while similar vectors could be found far from each other. This could result in 
two different populations of galaxies overlapping in the resulting SOM, or with
a single population being artificially split between two or more regions in the
map. It has been suggested that the first problem could be overcome by running 
the SOM several times, each time starting with a different set of random 
characteristic vectors, and choosing only the ``best'' run, e.g., in the sense
of minimizing the $\chi^2$ difference between all objects and the 
characteristic vectors of the nodes to which they were mapped. However, since 
this is not a supervised learning process, there may be many very different 
minima of this measure, each corresponding to a different topology, with little
to choose between them. In addition, this solution does not answer the second 
problem we raise. Furthermore, randomizing the initial characteristic vectors 
makes the entire process unrepeatable. 

In order to avoid these difficulties we first run a simple clustering algorithm
(SCA) on the data, and use the emerging crude clusters to decide how to 
initialize the map vectors. Our version of the SOM algorithm consists of two 
stages : in the startup phase we employ the SCA to get a rough idea of how the 
objects cluster. The SCA initially defines each object in the dataset as an 
independent ``group'' in data-space. The $L2$ norm (Euclidean distance) is 
adopted as the data-distance measure, and a search radius is defined, which 
increases linearly with the number of iterations. In each iteration groups 
whose centers of mass lie within the search radius of each other are merged, 
and so the number of groups decreases monotonically with time. The stopping 
criterion for the SCA is met once the three largest groups contain between them
more than half of the vectors. The number was set to three because three 
vectors define a plane and can therefore be mapped in a topologically-faithful
manner onto our two-dimensional map. Note, however, that the three largest 
groups need not represent the most diverse combinations of the morphological 
parameters. For this reason we examine all groups containing more than $1\%$ of
the data when the SCA is stopped (typically of order ten groups). Out of all 
the vectors representing the ``centers of mass'' of these groups, we select 
those which contain a maximum or a minimum value of at least one of the 
parameters. Since we are using four parameters, the number of such selected 
vectors ($N_v$) is in the range $[2,8]$, but expected to be closer to $8$ in 
most cases. The results of running the SCA (and any other crude clustering 
algorithm) over a given dataset are expected to be quite independent of the 
exact details of the algorithm. Different distance measures may result in 
somewhat different results but since we are using the SCA only as the first 
stage in our analysis, such differences are not important for the final
outcome.

In the second stage we iterate through all possibilities of selecting three
so-called ``anchors'' out of these $N_v$ vectors to initialize and train the 
SOM. The selected vectors are assigned to three nodes in the map in a way that 
conserves their relative data-distances. All other nodes {\it within the 
triangle enclosed by these anchors} are then assigned characteristic vectors
that are weighted averages of these three key vectors, the weight being the 
inverse of the map-distance from each anchor-node :

\bigskip
\begin{math}
$$
(1)~~~~~~~~~~~~~~~~~{\bf C}^{(i,j)} = \frac {\sum_{k=1}^3 {\bf C}_k / 
                            d_k^{(i,j)}} {\sum_{k=1}^3 1/d_k^{(i,j)}}
$$
\end{math}
\bigskip

\noindent where for $k \in \{1,2,3\}, {\bf C}_k$ is one of the three key
characteristic vectors and $d_k^{(i,j)}$ is the map-distance between node 
$(i,j)$ and the node in which that key vector resides. Only the region inside
the triangle is used. This procedure allows the map nodes to span much of the
variance in the data from the outset, and guarantees the repeatability of the
results. We select only three anchors because, again, three points define a
plane and can therefore be mapped in topologically-faithful manner onto the
two dimensional map. Choosing all possible combinations of three vectors for 
the role of anchors allows us to search for the combination that best 
represents the data, in an unsupervised way. Repeatability is guaranteed
because the entire process is deterministic and does not require input from 
the user.

Next comes self-organization. We again adopt the $L2$ norm as our 
data-distance measure. For each data vector ${\bf V}$ (describing one galaxy)
the winning node is node $(i,j)$ for which the data-distance between its 
characteristic vector ${\bf C}^{(i,j)}$ and the data vector ${\bf V}$ is 
minimal. This distance is given by :

\bigskip
\begin{math}
$$
(2)~~~~~~~~~~~~~~~~~d(i,j) = \displaystyle{\surd \sum_{l=1}^M (C^{(i,j)}(l) - 
                              V(l))^2 }
$$
\end{math}
\bigskip

Once the entire dataset has been mapped to the SOM the characteristic vectors
of every node are updated. There are two possible sources of contribution to 
the update of a given node : one is due to all the objects that were mapped 
directly onto that node; and the other is due to the objects mapped to nearby 
nodes, which affect that node by virtue of the attempt to conserve topology. 
Let $\langle {\bf V}^{(i,j)} \rangle$ be the average of all vectors mapped onto
node $(i,j)$. Then the first contribution is of the form :

\bigskip
\begin{math}
$$
(3)~~~~~~~~~~~~~~~~~{\bf dC_1}^{(i,j)} = \langle {\bf V}^{(i,j)} \rangle
$$
\end{math}
\bigskip

\noindent and the contributions of the second kind will come from nodes 
$(i_1,j_1)$ around node $(i,j)$ and will each have the form : 

\bigskip
\begin{math}
$$
(4)~~~~~~~~~~~~~~~~~{\bf dC_2}^{(i,j) | (i_1,j_1)} = \exp[{-(d_m^{(i_1,j_1)})^2
                                   / 2\sigma^2}]~ \cdot ~\langle 
                                   {\bf V}^{(i_1,j_1)} \rangle
$$
\end{math}
\bigskip

\noindent where $d_m^{(i_1,j_1)}$ is the map-distance between $(i,j)$ and 
$(i_1,j_1)$. The ``environment kernel'' chosen here is a Gaussian whose width,
$\sigma$, is a decreasing function of the number of iterations $n_i$ :

\bigskip
\begin{math}
$$
(5)~~~~~~~~~~~~~~~~~\sigma (n_i) = \frac {1}{n_i}
$$
\end{math}
\bigskip

The reason for the dependence of $\sigma$ on the number of iterations is that 
as the structure of the map becomes more organized it is desirable to limit 
the effect of the environment. If the other nodes were always allowed to 
contribute at the same level the process of self-organization might never 
converge and finer details in the map could be washed away. For practical 
purposes of reducing the number of calculations, the environment of node 
$(i,j)$ from which the nodes $(i_1,j_1)$ are taken is limited to a square of 
side 7 (i.e., vertical/horizontal map-distance of no more than 3) centered on 
node $(i,j)$. There is no need to go any further, because even when $\sigma$ is
maximal at 1 (during the first iteration), the coefficient $d_m$ drops to about
$0.01$ at a map-distance of 3, and therefore nodes further away from $(i,j)$ 
are unlikely to contribute to it significantly. The updated value of the
characteristic vector of node $(i,j)$ is therefore given by :

\bigskip
\begin{math}
$$
(6)~~~~~~~~~~~~~~~~~{\bf C}_{new}^{(i,j)} = (1-\eta) \cdot 
   {\bf C}_{old}^{(i,j)} + \eta \cdot \frac {\displaystyle{{\bf dC_1}^{(i,j)} 
   + \sum_{(i_1,j_1) \neq (i,j)} {\bf dC_2}^{(i,j) | (i_1,j_1)}}}
   {\displaystyle{1 + \sum_{(i_1,j_1) \neq (i,j)} \exp[-(d_m^{(i_1,j_1)})^2 / 
    2\sigma^2]}}
$$
\end{math}
\bigskip

\noindent where the denominator in the second term is the normalization of all 
the weighted contributions, and $\eta$ is a parameter which describes the 
``learning rate'' of the SOM. We set $\eta$ to $0.02$. It is not advised to 
make $\eta$ large because then the changes in the characteristic vectors can 
become erratic.

At the end of each iteration we monitor the root mean square difference between
the current and previous characteristic vector of each node. We stop training 
the map when the largest of these differences has dropped below $0.1\%$ of its
maximal possible value. Typically this leads to convergence within several
thousand iterations.

Self organization is repeated for all selections of three anchors. For each
such selection all the vectors in the dataset are mapped onto the trained SOM
and the $\chi^2$ difference between the data vectors and the characteristic 
vectors of the nodes to which they were mapped is monitored to find the best 
triplet. The SOM resulting from the best triplet is then chosen as the best
overall SOM.

\section{An Example : Non-Linear Mapping in Four Dimensions}

We test the ability of the SOM to handle non-linear mapping in multi-dimensions
by first defining a curve in a space of the same dimensionality as our galaxy 
dataset. In order to demonstrate the ability of the SOM to retain topological 
information, the curve is specified in parametric form. This conveys a clear 
notion of the order of points along the curve. The curve is given by :

\bigskip
\begin{math}
$$
(7)~~~~~~~~~~~~~~~~~{\bf F}(\theta) = (\sin (\theta), \cos (\theta), \sin 
(\theta) \cos (\theta), \sin^2 (\theta))
$$
\end{math}
\bigskip

\noindent where $\theta$ is the free parameter. We choose five points along the
curve, corresponding to $\theta$ values of $\pi/12, \pi/6, \pi/4, \pi/3$ and 
$5\pi/12$. Around each point we randomly scatter 400 points. There is little 
overlap between these five clouds of points and the relations between any two 
of the components of ${\bf F}$ are all non-linear. The SOM software is trained 
on a dataset containing all 2000 points and the results are shown in figure 1. 
The top left panel shows the mapping of the full dataset, and there appear to 
be three to five distinct concentrations. The other five panels each depict one
group of points (denoted by the corresponding value of $\theta$). It is plain 
to see that the SOM {\it maintains the order of the groups along the curve}, 
although some mixing between adjacent groups takes place. The SOM is therefore
capable of mapping non-linear datasets while conserving much of the topology. 
Note also that although the initialization of the SOM is in the form of a 
triangle, in this case it is an obtuse angle triangle, closely resembling the 
true shape of the distribution of points in the original space - that of a
one dimensional curve in a four dimensional space. One possible drawback of 
this representation is that close to the vertices of the triangular region
groups tend to be more concentrated, implying a steeper gradient in the 
parameters associated with vectors near the vertices. The mapping is therefore 
not totally faithful topologically. As mentioned above, the numbers along the 
axes represent positions in the map, not values of $\theta$ or any other 
parameters. The different panels of figure 1 show the dependence of $\theta$ on
position in the map.

\section{Galaxy Distributions in Morphology Space}

As a preliminary step Principal Components Analysis (PCA) of the dataset was
performed, which seeks to represent as much of the variance in the data by
replacing the original axes by {\it linear} combinations of them. However, the
first PC only spans $48\%$ of the variance and the first two PCs span only
$71\%$ of it. PCA is therefore inadequate for mapping these data in two
dimensions and a non-linear method is indeed required.

\subsection{Mapping Galaxy Populations}

We next proceeded to analyze the sample of 2934 galaxies with the SOM software.
The best resulting map (in terms of $\chi^2$ between the data and the nodes to 
which they were mapped) is shown in figure 2. Shading progresses from light for
low population levels to dark for highly populated regions. Although only the 
vertices of the triangular map were initialized with vectors corresponding to 
actual clusters of data points, the final map is well populated in all nodes. 
This shows that the SOM training process refines the crude results of the 
clustering algorithm and brings out finer structure. However, mapping the full 
dataset like this is not very informative without examining the characteristic
vectors associated with each node. In figure 3 we show four panels, each 
depicting the distribution of values of a single morphological parameter in 
the SOM. There are apparent trends in the parameter distributions : blobbiness 
is lowest around the left vertex, and grows as one moves right, especially 
towards the upper right. Center displacement is highest in the top right vertex
and decreases toward both of the other vertices. The filling factor generally 
grows along the same direction but then decreases again. The skeleton ratio has
the clearest trend, growing strongly as one moves away from the bottom right 
vertex. 

With the help of figure 3 one can now identify the morphologies associated with
the map of figure 2 : the area of the left vertex is populated by smooth, 
symmetric galaxies with a high filling factor. This description corresponds to 
the appearance of elliptical galaxies. As one moves right towards the center of
the map, two trends become apparent : towards the top right vertex galaxies are
much more blobby with increasing asymmetry (higher center displacement). The 
filling factor drops but the skeleton ratio is high, so this region should 
correspond to images with a lot of elongated structure, such as spiral galaxies
or galaxies with tidal tails. Towards the bottom right the skeleton ratio drops
sharply while the values of the filling factor and the center displacement do 
not have a clear trend. This implies galaxies of generally ``knotty'' 
appearance, some of which are very asymmetric with a lot of apparent structure,
while others are less asymmetric and exhibit less structure. These morphologies
largely correspond to peculiar galaxies. 

In order to verify the above interpretations and to study how different 
properties of galaxies correspond to their morphology we defined subsets of our
sample according to several criteria, and mapped these subsets onto the trained
SOM. In figure 4 each panel shows the mapping of one subset, {\it normalized to
the total size of that subset}. This means that the intensities are relative 
within each panel and should not be directly compared between panels. The 
panels in the bottom row depict populations selected by eyeball classification.
Such classifications were made by one of us (AN, see Naim {\it et al.} 1997) 
for roughly one third of the entire dataset, as a preparation to using {\it
supervised learning} for these galaxies. Ellipticals and S0's are depicted in 
the left panel, spirals in the middle panel and peculiars in the right panel. 
The locations of these subsets on the map match what we expect from the 
analysis of the characteristic vectors above : ellipticals are mostly 
concentrated around the left vertex; spirals are well spread out but appear 
more concentrated towards the center of the map; peculiars are mostly found in 
the right hand side of the map. This morphological sequence generalizes the
one dimensional Hubble Sequence into a two dimensional map. Roughly speaking, 
the horizontal axis depicts mostly the change in overall smoothness and 
symmetry of images, while the vertical axis describes the nature and frequency
of structure in the images. 

The top three panels depict the distributions of galaxies in three subsets 
selected by apparent isophotal magnitude. The left panel depicts galaxies 
brighter than $I=21$, the middle panel depicts galaxies in the range $22<I<23$,
and the right panel contains galaxies fainter than $I=23.5$. The gaps in the
ranges of apparent magnitude shown in these panels are intended to reduce the 
overlap and accentuate trends, since the distributions form a continuum. The 
same applies to the panels describing color and bulge dominance, below. 
However, these magnitude ranges were chosen a-priori. The magnitude limits 
represent a compromise between representing the full range of magnitudes and
ensuring that no single bin is underpopulated. There are 278 galaxies brighter 
than $I = 21$, 722 in the range $22 < I < 23$ and 946 fainter than $I = 23.5$.
The shift in the concentrations of galaxies with apparent magnitude is evident.
Since the redshift distribution of galaxies is a function of apparent magnitude
these three panels may describe in a statistical way the evolution of galaxy
morphologies with redshift. Verifying this would require many spectroscopic 
redshifts, though, and work is in progress along these lines (Naim {\it et al.,
in preparation}). The trend we see here is clear : at the bright end the 
smooth, symmetric galaxies are much more prominent than at the faint end. 

The panels in the second row depict subsets selected according to the only
available color, $V-I$. The left panel contains red galaxies, with (isophotal)
$V-I > 1.8$. The middle panel contains intermediate color galaxies ($1.0 < V-I
< 1.4$), and the right panel depicts blue galaxies ($V-I < 0.6$). Color 
appears to follow morphology, albeit with significant scatter. There is a 
preference of blue galaxies to occupy the upper half of the right side of the 
map. The third row panels describe subsets selected by Bulge-to-Total ratio, 
defined as the light contribution of the bulge component over the combined 
contributions of the bulge and disk components. This ratio is calculated from 
the maximum-likelihood fits of bulge and/or disk models to the galaxy image 
(Ratnatunga {\it et al.} 1997 contains many details about the subtleties of 
these fits). The left panel describes bulge dominated galaxies ($B/T > 0.8$), 
the middle panel describes intermediate cases ($0.3 < B/T < 0.7$) and the right
panel depicts disk dominated galaxies ($B/T < 0.2$). Interestingly, the 
bulge-dominated galaxies appear less concentrated in the right hand side than 
the intermediate cases. We verify this impression in figure 5, where we show
the mean positions of five subsets, selected by $B/T$ ratios, on the trained 
SOM. The scatter around these means is considerable, but there is nevertheless
a general trend of moving leftwards with increasing $B/T$ ratio, which is
reversed by the last subset. This is an indication of a change in morphology 
among bulge-dominated galaxies. Closer examination of figure 2 confirms that 
galaxies with $B/T > 0.8$ cluster in two regions, one corresponding to smooth, 
symmetric morphologies and one corresponding to blobby and asymmetric 
morphologies. This latter population has already been noted (Naim {\it et al.} 
1997). It may correspond to the ``blue nucleated galaxies'' found in the 
Canada-France redshift survey (Schade {\it et al.} 1995), although verifying 
this point requires further work. 

\subsection{The Effect of Noise}

One possible source of the apparent correlation between blobbiness and 
asymmetry of images on the one hand and their apparent magnitude on the other,
is that as one looks at fainter images noise sets in and changes their 
appearance. In order to check this we show in figure 6 how galaxies of high 
integrated signal-to-noise index are mapped on the same trained SOM. While the 
full sample contains galaxies that virtually all have $\nu > 100$, the subset 
shown in figure 6 was selected to have $\nu > 500$. The fraction of bright 
galaxies in this subset is naturally higher than in the full sample, so it is 
difficult to completely decouple the effect of reducing the noise from that of 
selecting brighter galaxies. Nevertheless, while the concentrations of blobby, 
asymmetric galaxies appears less prominent in figure 6, it is still a 
significant population. Had that population been due only to noise it should 
have disappeared in this figure completely. We thus conclude that blobby,
asymmetric galaxies indeed exist and their numbers do increase as one goes to 
fainter magnitudes.

We turn back to figure 3 now, in order to see how noise might have affected the
evaluation of our parameters. The panels describing the distributions of
blobbiness, isophotal filling factor and isophotal center displacement 
show the trends one would expect, although finer details are also evident,
allowing one or more parameters to vary slightly from one node to the next.
The one problematic parameter is the skeleton ratio : while the map shows the
expected small values in the region of the peculiars (due to nearly round 
star-forming regions) and higher values in the region occupied by the spirals
(due to elongated arms), the values are disturbingly high for ellipticals, for
which one would expect no features at all (and consequently a value of zero
for the skeleton ratio). We note that unlike the other three parameters which
were evaluated from the raw images of the galaxies, the skeleton ratio is 
measured from the residual images, left after the best fitted photometric model
had been subtracted. The skeleton ratio is measured for features that stand out
relative to the residual image, and when the residual contains no real features
(e.g., in an elliptical) noise may result in the ``detection'' of spurious 
structure. We suspect that this is the source of the relatively high skeleton 
ratio which characterizes nodes in the region occupied by ellipticals, but 
further work is needed in order to verify that these features are not real. 
Luckily, this effect appears to influence most of the bulge-dominated, 
featureless galaxies in the same way, thus not disturbing their clustering 
properties. On the other hand, the skeleton ratio is very useful in 
distinguishing spirals from peculiars, and should not be discarded.

\section{Discussion}
 
It has always been important to examine individual galaxies in detail and study
the processes dictating their appearance. However, for the fuller picture of 
galaxy evolution one requires statistical analysis. Quantitative parameters are
required which capture the diversity of galaxy morphologies, while not becoming
too specialized or numerous. Here we continue to use the set of four parameters
introduced in a previous paper (Naim {\it et al.} 1997). However, unlike that
paper, our aim here is to analyze these data in an unsupervised way, in order 
to learn new things. One serious difficulty that arises with even a modest 
number of parameters is visualizing data in more than three dimensions. We 
therefore make use of our variant of the Kohonen Self Organizing Map (SOM), 
which allows one to cast a distribution in multi-dimensions onto a two 
dimensional map. Our algorithm is not necessarily the best for this purpose 
and other algorithms exist. Using SOMs allows us to visualize the distributions
of galaxies and point out interesting populations for further study. In this 
respect the SOM is a diagnostic tool, facilitating the first step that needs to
be taken with any kind of data : looking at them.

We examine the SOM on a synthetic dataset and confirm its ability perform 
non-linear mapping while maintaining the correct topological order of the
higher dimension space. We then apply it to our HST galaxy sample.
In the resulting SOM galaxies cluster in several groups in morphology space. 
We confirm the picture which emerged from previous work (Glazebrook {\it et 
al.} 1995; Driver {\it et al.} 1995; Abraham {\it et al.} 1996), according to 
which the galaxy population becomes more and more dominated by blobby, 
asymmetric morphologies as one goes fainter. Further, we show that the colors
of galaxies at moderate redshifts become significantly bluer. This could be 
partly due to the shift in rest-frame band as one goes to higher redshifts, but
actual measured redshifts are needed in order to evaluate how much this effect 
contributes to the trend we see in the SOM. Bulge dominance also appears 
related to morphology, the blobby galaxies being more disk dominated. However, 
a population of bulge-dominated galaxies with high blobbiness and asymmetry,
which has been noted by supervised classification (Naim {\it et al.} 1997), is
rediscovered here in an independent way. Note that this is achieved without
any recourse to eyeball classification, i.e., its existence can be inferred 
without suspecting it from the outset. Bulge dominated galaxies with blue
colors were also found by Koo {\it et al.} (1995), and some of them exhibit
peculiar morphologies (e.g., ``knots''). That study was limited to a small
number of galaxies and therefore no statistical conclusions can be drawn
regarding the bulge dominated peculiars. Pascarelle {\it et al.} (1996) report
the discovery of compact (half light radius around $0.1$ arcsec), blue objects
which are apparently sub-galactic clumps. It is possible that these clumps,
once assembled more closely, give rise to the bulge dominated peculiars that
we identify in our sample, although this is by no means certain. Alternatively,
bulge-dominated peculiars may be older galaxies caught in the process of 
merging with dwarf companions. We have no way of telling with current data.

Noise becomes progressively more important as one goes to fainter images, but
our analysis shows that it can not fully account for the trends we detect. The
skeleton ratio parameter is most affected by noise in smooth, symmetric images
but does not significantly bias the clustering properties of that population as
a whole, and is still very useful in separating two other populations 
(corresponding to eyeball classes of spirals and peculiars). An improved
version of this parameter may give better results, though. K-corrections are
also of great importance at redshifts of order unity, as discussed, e.g., by
Odewahn {\it et al.} (1996). However, we have not studied their effect on our 
parameters in this paper, because we only have two filters for the data 
presented here (I and V). A study into the effect of the filters used on the 
measured morphological parameters is currently under way, using MDS fields that
were taken in three filters (B, V and I). Any effect the K-corrections may 
have on our parameters will, of course, influence the resulting SOM. 

To summarize, since morphological classification has become too refined we 
adopt an approach which utilizes morphology without any classification. The 
SOM succeeds in mapping different morphologies to different regions of the map,
and we are encouraged by the apparent correlation of morphology with other 
quantities, such as color and bulge dominance. These correlations allow us to 
use morphology as a selection criterion for further studies of specific 
populations (e.g., mergers). However, understanding galaxy evolution requires 
the addition of more physical information, such as rotation curves and full 
spectral analysis. In this paper we propose a framework into which such 
information could be incorporated once it becomes available. Our hope is that 
this modest first step will eventually lead to the emergence of an overall 
scheme incorporating most aspects of galaxy formation and evolution.

\bigskip

{\bf Acknowledgements}

We would like to thank Ofer Lahav, Jens Hjorth, Bob Abraham and Richard Ellis 
for raising important points regarding SOMs and morphology and its 
implications. As always, Eric Ostrander's contribution to the MDS pipeline was
invaluable. We also thank the referee for a thorough report.
This research was supported by funding from the HST Medium Deep 
Survey under GO grants p2684 {\it et seq.}

\bigskip

{\bf Figure captions}

\bigskip

{\bf Figure 1} : The Mapping of the Synthetic Dataset onto its SOM. Numbers 
along the axes represent position in the map, not values of any of the four
dimensions of the data.

\bigskip

{\bf Figure 2} : The Mapping of the Full Sample (2934 Galaxies) onto its
Trained SOM. The darker the shade, the more populated the node is. Again,
numbers along the axes denote position in the map, not values of morphological
parameters.

\bigskip

{\bf Figure 3} : The Distributions of Parameter Values in the Trained SOM. Top
left : Blobbiness; top right : Isophotal Filling Factor; bottom left : 
Isophotal Center Displacement; bottom right : Skeleton Ratio. The darker the
shade the higher the value of the parameter.

\bigskip

{\bf Figure 4} : The Mapping of Subsets of the Sample onto the Trained SOM. Top
row : subsets selected by Apparent I Magnitude; second row : by Color; third 
row : by Bulge Dominance; bottom row : by Eyeball Classifications. Refer to 
figure 3 for the changes in each of the four parameters as a function of
position in the map.

\bigskip

{\bf Figure 5} : The mean positions in trained SOM of subsets selected by
$B/T$ ratio. The scale is the same as the one in figure 2. The trend set by 
the subsets up to $B/T$ of $0.8$ is reversed by the $0.8 < B/T \le 1$ subset, 
indicating the existence of bulge dominated galaxies with blobby, asymmetric 
morphologies. Refer to figure 3 for the changes in each of the four parameters 
as a function of position in the map.

\bigskip

{\bf Figure 6} : The Mapping of High Signal-to-Noise images ($\nu > 500$) onto 
the Trained SOM. Compare with figure 2.

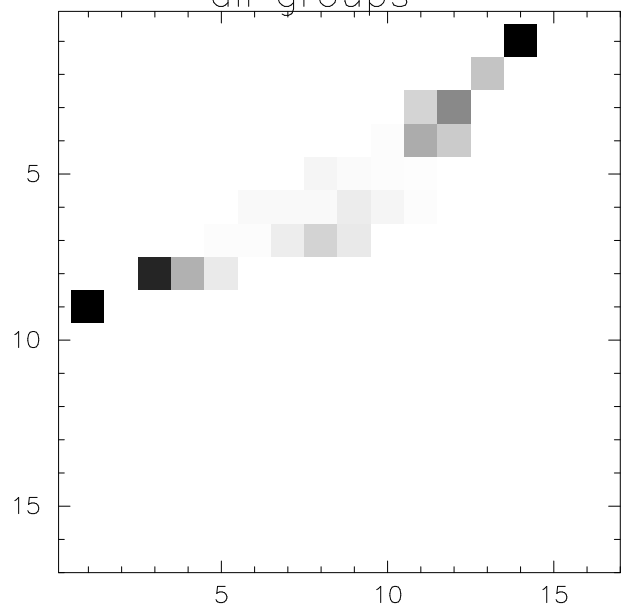
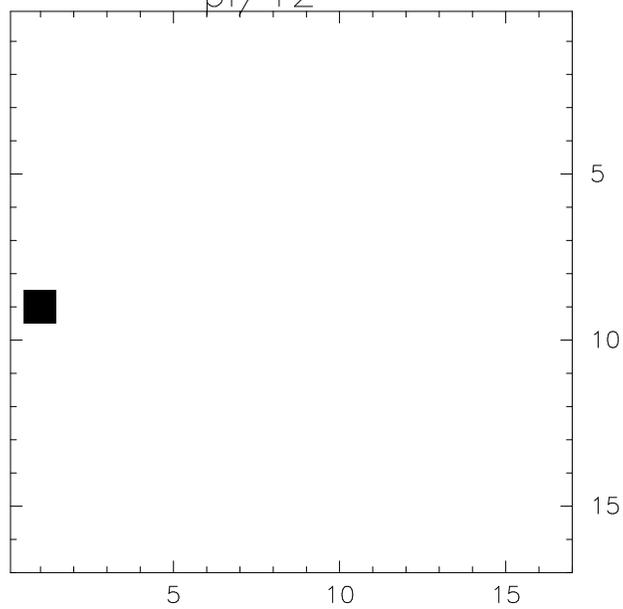
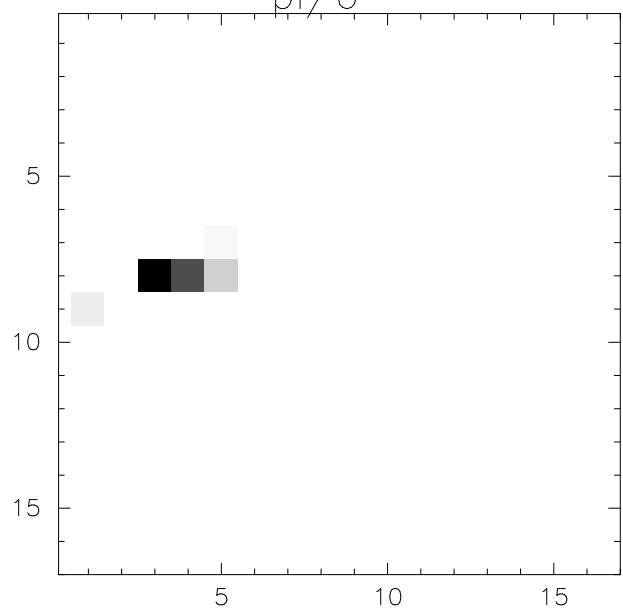
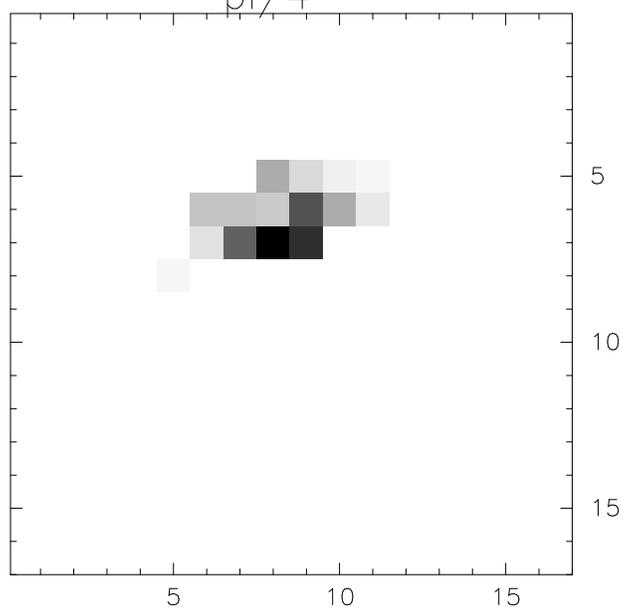
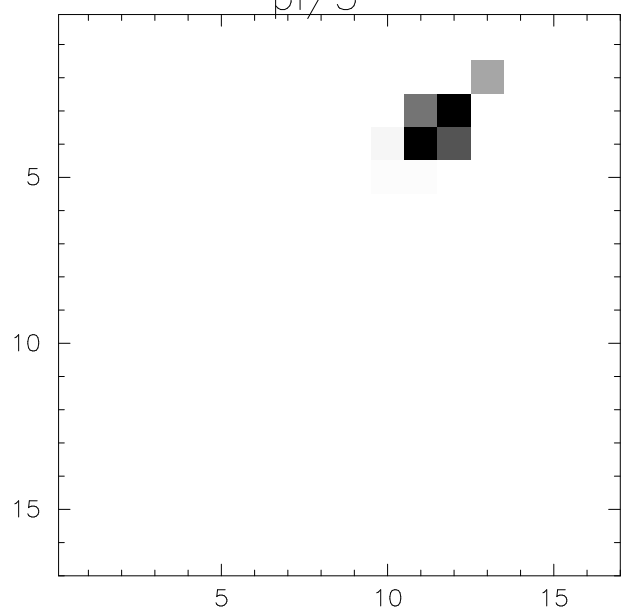
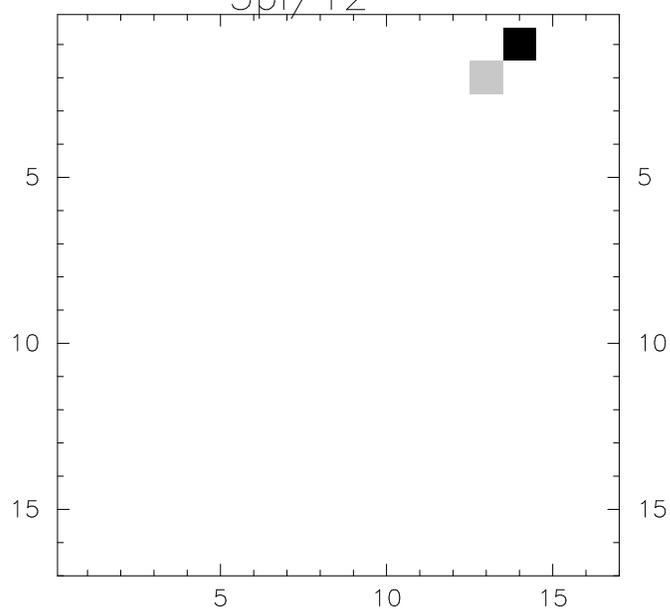
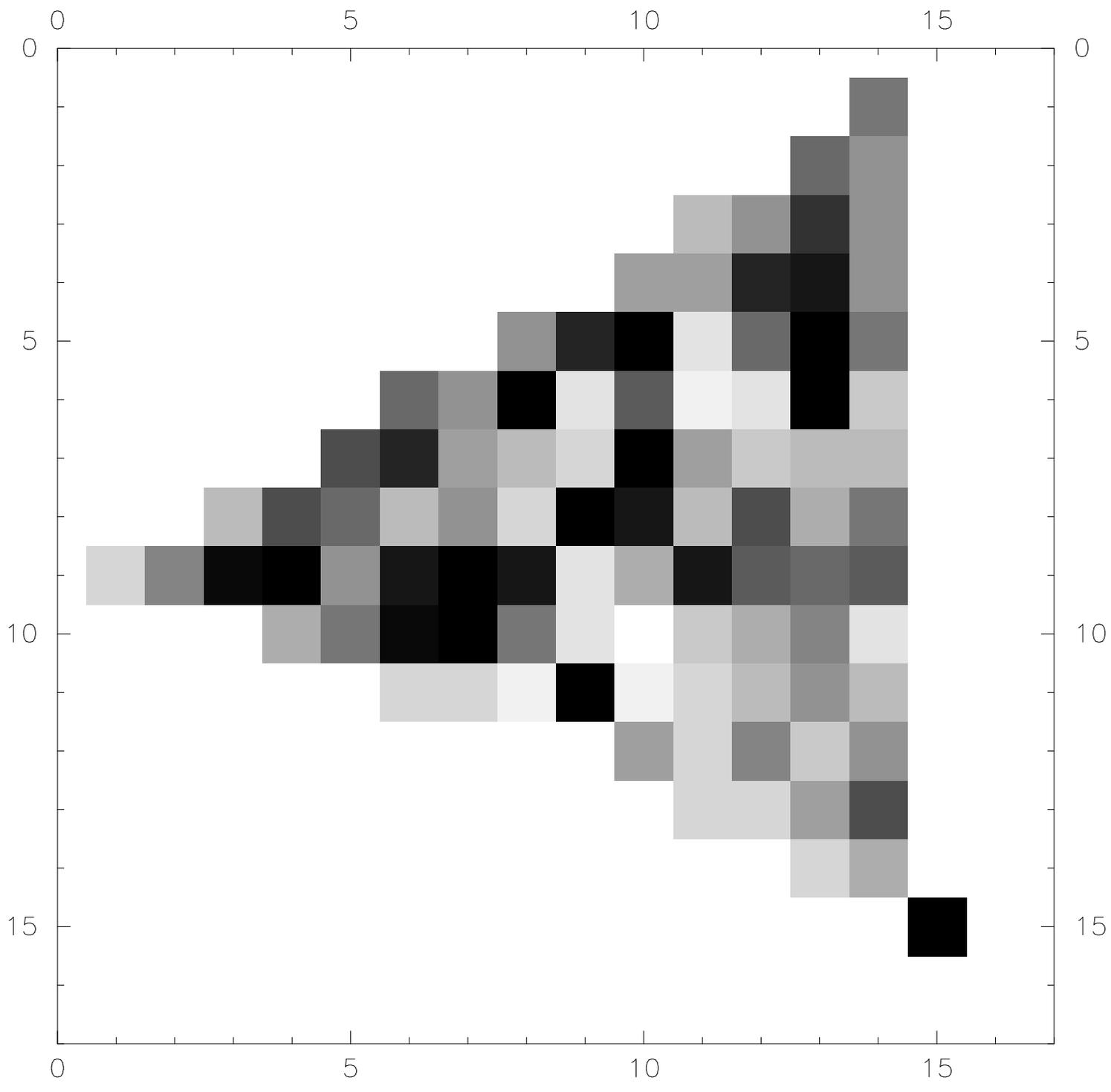
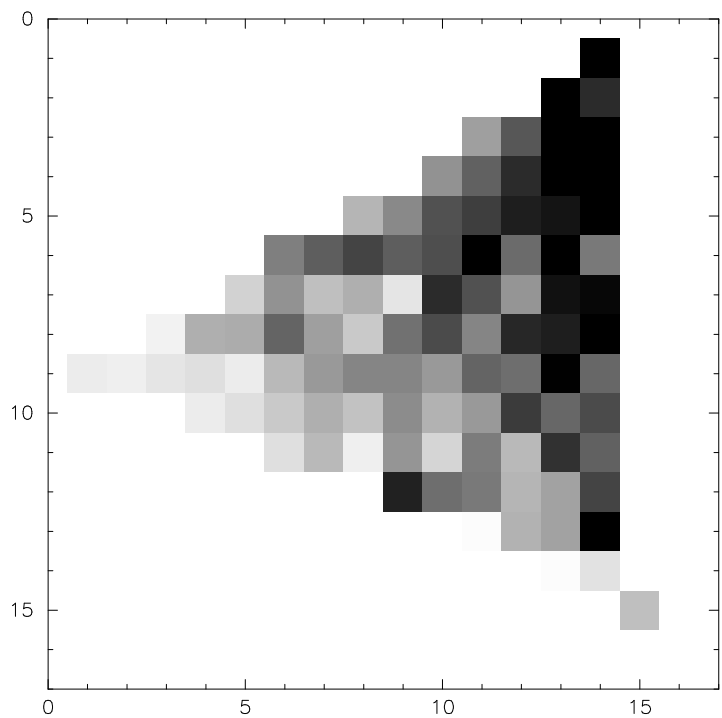
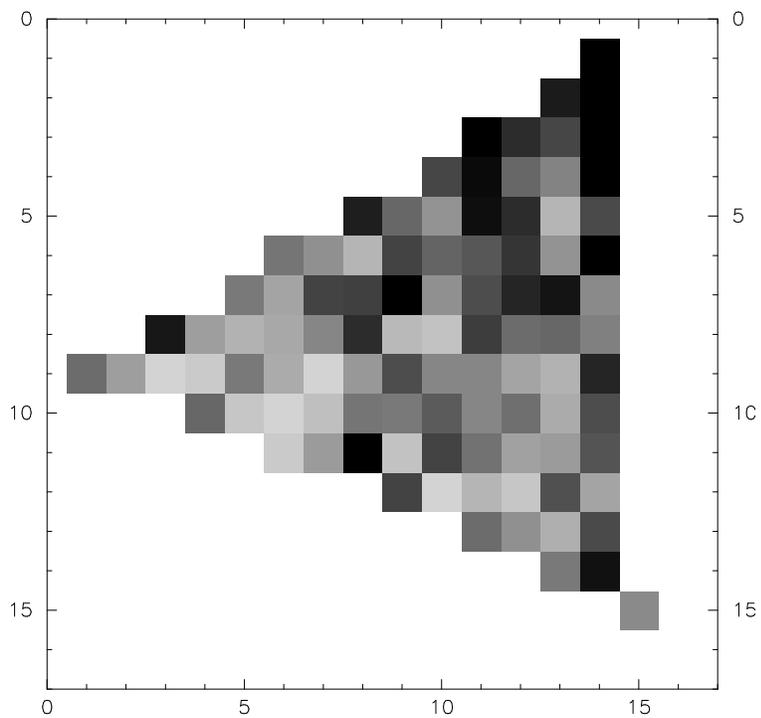
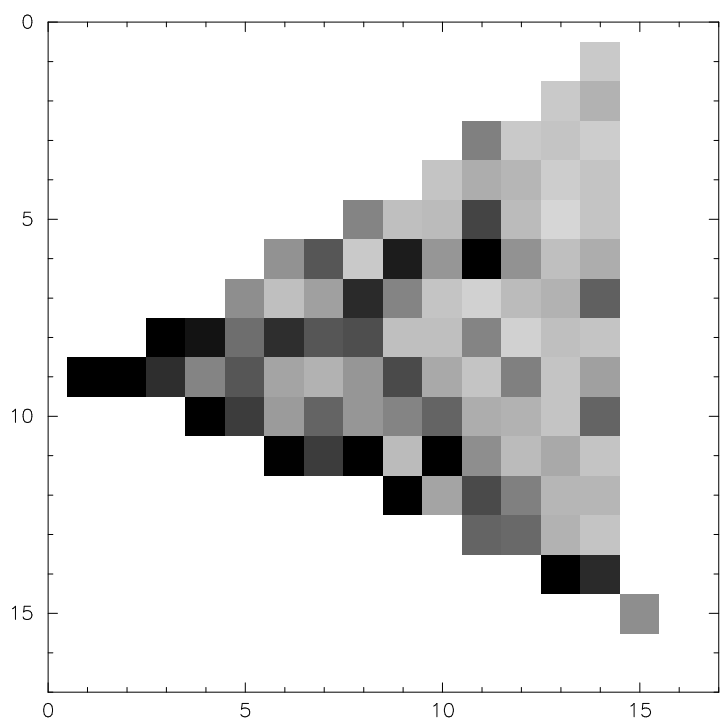
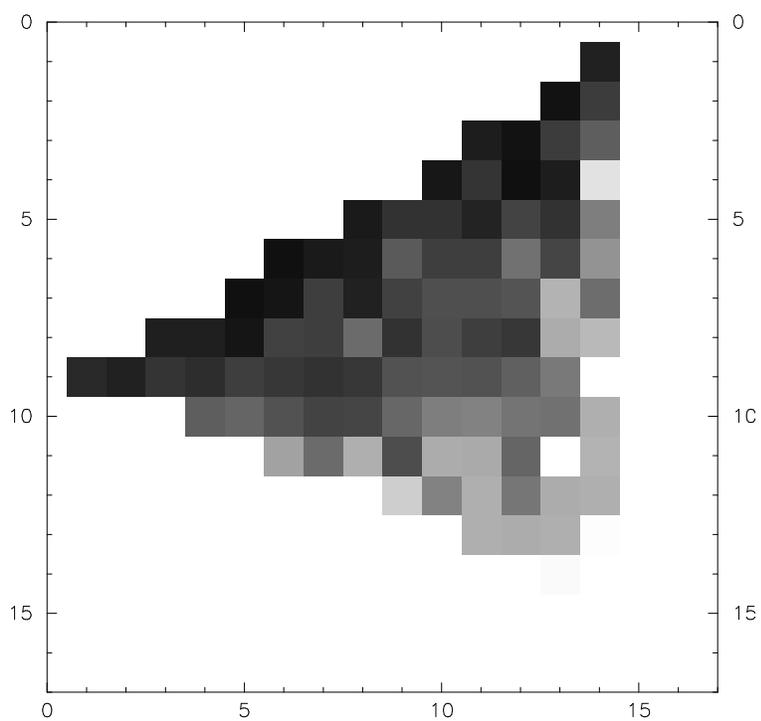
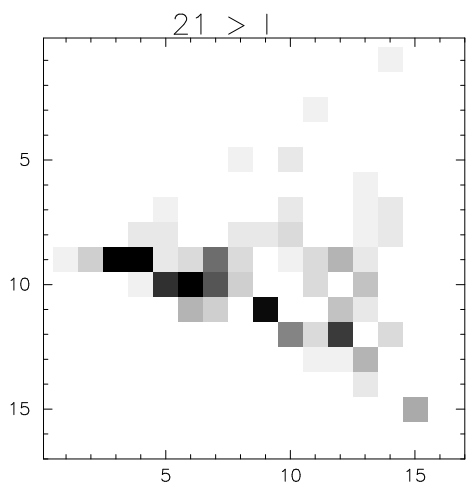
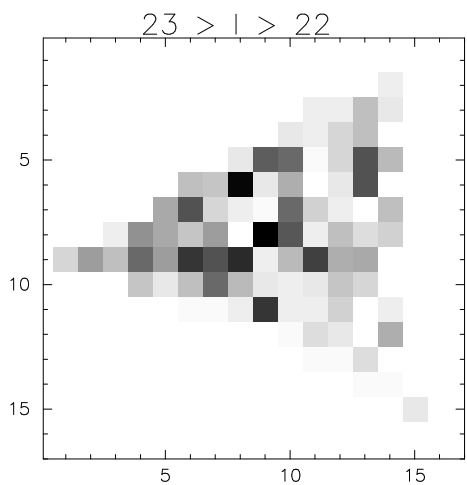
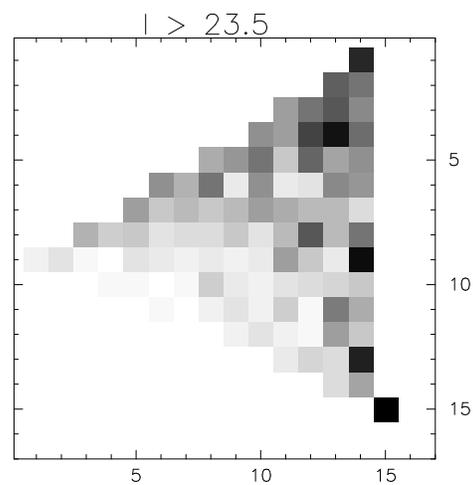
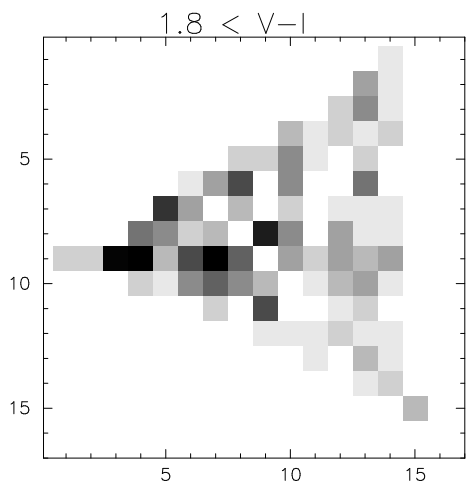
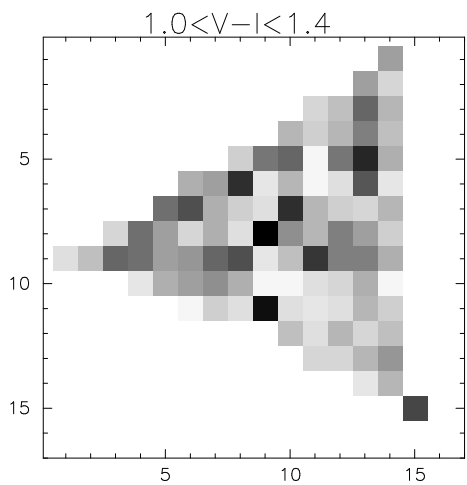
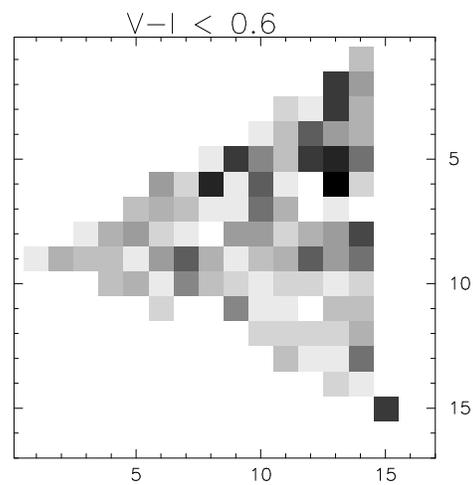
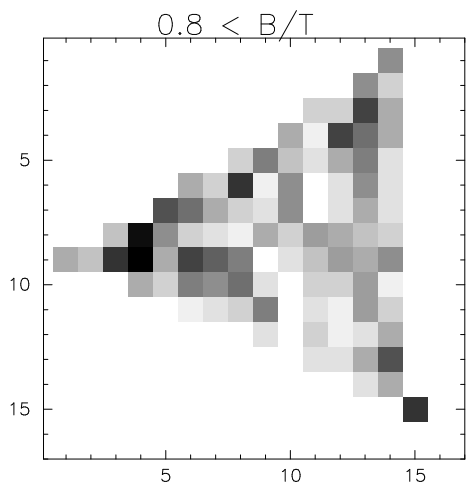
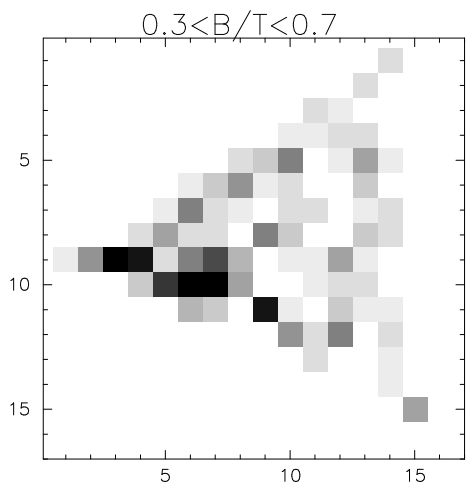
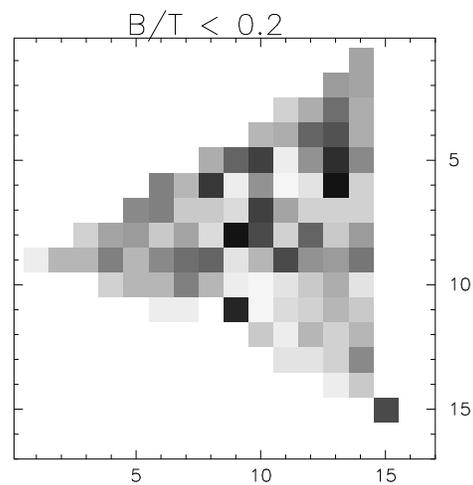
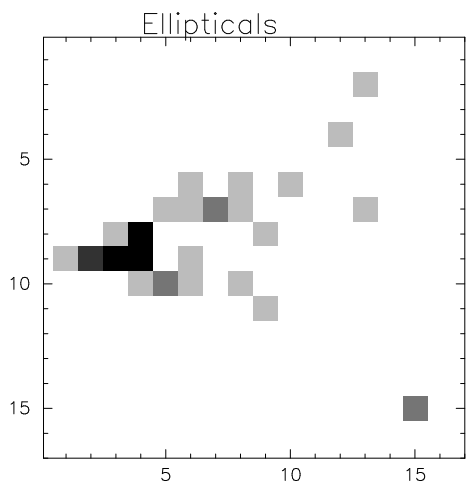
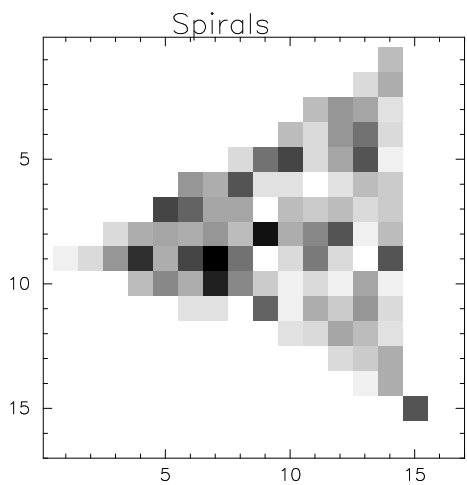
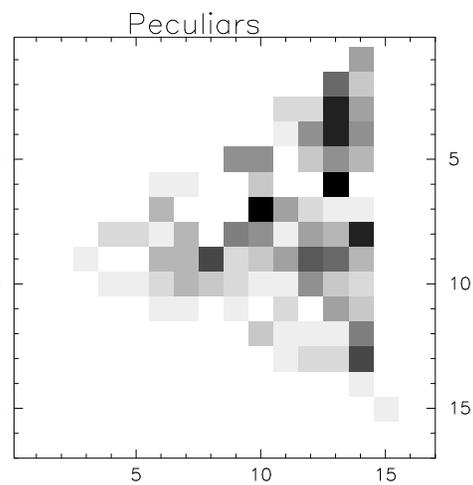
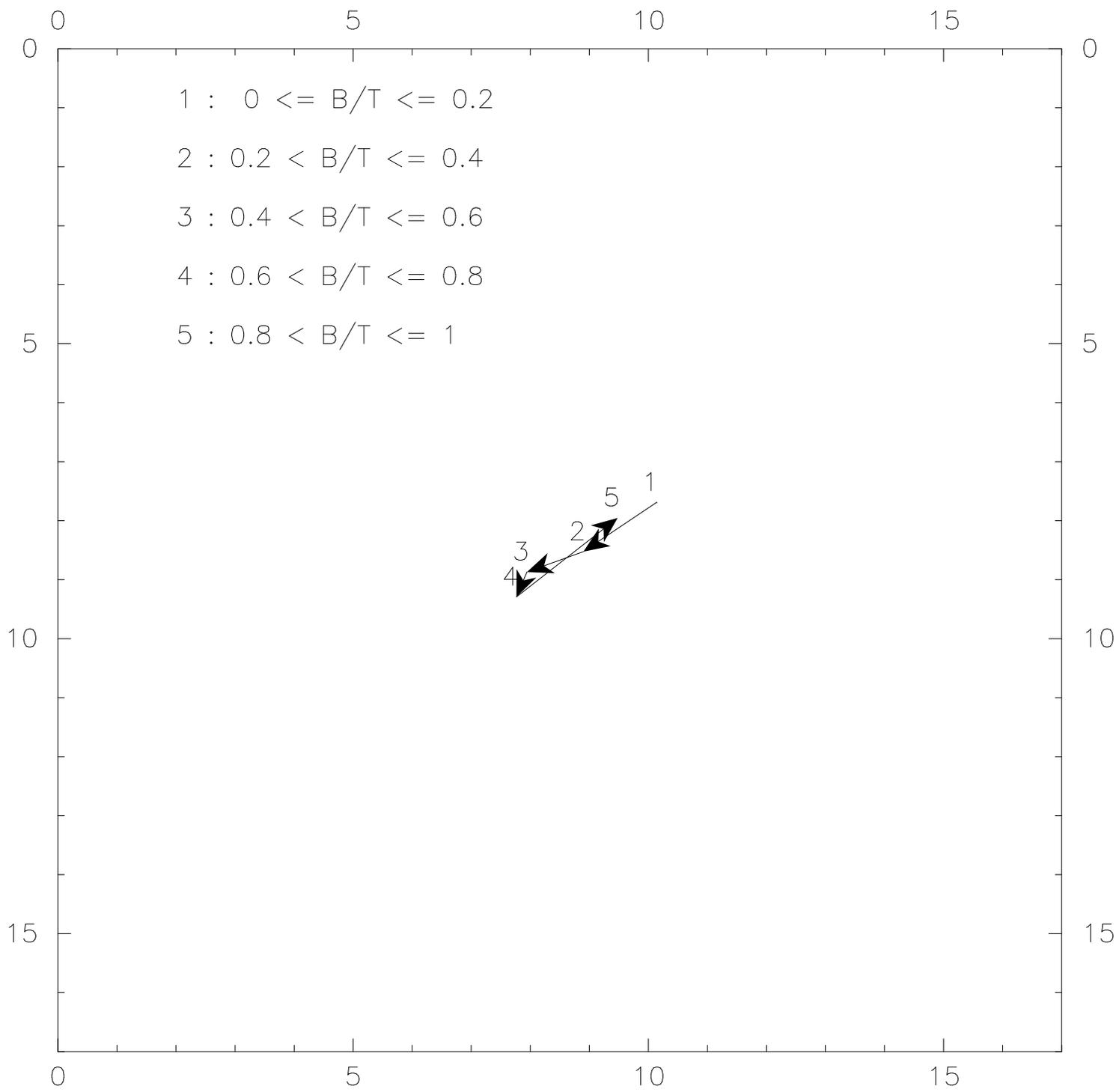
\end{document}